\documentclass[aps,longbibliography,superscriptaddress,10pt]{revtex4-1}

\usepackage{graphicx}
\usepackage{color}
\usepackage{subfigure}
\usepackage{amsmath}
\usepackage{amssymb} 
\usepackage{color}

\usepackage[margin=1.25in]{geometry}

\begin{document}

\title{Deposition of a particle-laden film on the inner wall of a tube}

\author{Deok-Hoon Jeong}
\affiliation{Department of Mechanical Engineering, University of California, Santa Barbara, CA 93106, USA}

\author{Anezka Kvasnickova}
\affiliation{Department of Mechanical Engineering, University of California, Santa Barbara, CA 93106, USA}

\author{Jean-Baptiste Boutin}
\affiliation{Department of Mechanical Engineering, University of California, Santa Barbara, CA 93106, USA}

\author{David C\'ebron}
\affiliation{Universit\'e Grenoble Alpes, CNRS, ISTerre, Grenoble, France}

\author{Alban Sauret}\email{asauret@ucsb.edu}
\affiliation{Department of Mechanical Engineering, University of California, Santa Barbara, CA 93106, USA}

\begin{abstract}
The withdrawal of a liquid or the translation of a liquid slug in a capillary tube leads to the deposition of a thin film on the inner wall. When particles or contaminants are present in the liquid, they deposit and contaminate the tube if the liquid film is sufficiently thick. In this article, we experimentally investigate the condition under which particles are deposited during the air invasion in a capillary tube initially filled with a dilute suspension. We show that the entrainment of particles in the film is controlled by the ratio of the particle and the tube radii and the capillary number associated with the front velocity. We also develop a model which suggests optimal operating conditions to avoid contamination during withdrawal of a suspension. This deposition mechanism can also be leveraged in coating processes by controlling the deposition of particles on the inner walls of channels.
\end{abstract}

\maketitle

\section{Introduction}

Multiphase flows involving liquid phases, gas, and solid particles are ubiquitous in the dispensing of liquid, coating of tubings, and flows in porous media \cite{ingham1998transport,blunt2001flow,koch2001inertial,schwarzkopf2011multiphase}. The presence of particles in porous media is especially important since it can lead to clogging events \cite{dressaire2017clogging}. The liquid initially filling the tube or the pores is commonly displaced by a non-miscible fluid, \textit{e.g.}, air \cite{maxworthy1989experimental,strait2015two}. As an invading fluid pushes the other fluid forward in a tube, this latter leaves behind a thin liquid layer or a series of drops depending on fluid properties and velocity \cite{Aussillous2000,magniez2016dynamics,zhao2018forced,hayoun2018film}. The situation of a slug of wetting liquid pushed by an immiscible fluid has been considered in various studies, because of its relevance in liquid dispensing and coating of tubing \cite{taylor1961deposition,reinelt1985penetration,giavedoni1997axisymmetric,Aussillous2000}. When the gravitational and inertial forces are negligible, the thickness of the liquid film $h_0$ deposited on the inner wall of a capillary of radius $R$ by a Newtonian fluid of dynamic viscosity $\mu$ and surface tension $\gamma$ for a front velocity $U$ is governed by the competition between viscous and surface tension forces, captured through the capillary number ${\rm Ca}=\mu\,U/\gamma$ \cite[see, \textit{e.g.},][]{Aussillous2000,balestra2018viscous}. The film thickness $h_0$ in a capillary tube has initially been predicted by Bretherton in the limit ${\rm Ca} \ll 1$: ${h_0}/{R} = 1.34 {\rm Ca}^{2/3}$ \cite{bretherton1961motion}. The evolution of the film thickness for a broader range of parameters has been obtained through experiments and simulations \cite[see, \textit{e.g.},][]{balestra2018viscous}. In particular, the thickness of the film resulting from air invasion in a capillary filled with a Newtonian liquid is captured by the empirical relation \cite{Aussillous2000}:
 \begin{equation}\label{Taylor}
 \frac{h_0}{R} = \frac{1.34 {\rm Ca}^{2/3}}{1+3.35\,{\rm Ca}^{2/3}}.
 \end{equation}
 Different extensions of this model have considered curved \cite{muradoglu2007motion} and non-cylindrical channels \cite{hazel2002steady}. For a partially wetting fluid, the contact line motion also plays a role in determining the deposition patterns \cite{cueto2012macroscopic,zhao2018forced}.  Most past studies have considered one fluid pushed by a second immiscible fluid. The influence of particles on the deposition and the composition of the thin film coating the inner wall of the capillary remains elusive. This configuration is nevertheless relevant to model the flow of suspensions in porous media and contamination of tubings. A description of the thin film formation is required to account for the influence of the particles dispersed in the liquid. However, because the film thickness and the diameter of the particles can be of the same order of magnitude, the particles deform the air/liquid interface and may modify the interfacial dynamics. 

Previous works with particulate suspensions have shown that fluid-fluid interfaces can reduce the length scale of the flow to less than a few particle diameters \cite{bonnoit2010mesoscopic}. This situation occurs between two fluid-fluid interfaces, for instance, during the pinch-off of suspension, which is strongly modified by the presence of particles \cite{furbank2004experimental,bonnoit2012,chateau2018pinch}, and during the spreading and fragmentation of suspension sheets \cite{raux2020spreading}. When extruding a suspension from a nozzle, the ligament of liquid thins out and eventually pinches off to generate a droplet due to interfacial effects. In tubings and porous media, the situation becomes even more complex because of the presence of solid surfaces. This type of particle confinement is at play during the drainage of suspension films \cite{Buchanan:2007fs}, impact of a suspension drop on a substrate \cite{lubbers2014dense,grishaev2017impact}, the dip coating of suspensions \cite{xu2016particle,kim2017formation,luo2018particle,xu2019enhancement,kudrolli2020unstable}, viscous fingering \cite{xu2016particle,kim2017formation}, and bubble rise in confined suspensions \cite{madec2020puzzling}. The displacement of the particles is controlled by the capillary force, the drag force exerted by the fluid, and the friction on the solid substrate. The complexity of such a situation was recently considered by Yu \textit{et al.} \cite{Yu:2017kr}, who investigated the coating of the air/liquid interface of a long gas bubble translating in a mixture of glycerol and particles exhibiting a finite contact angle. Yu \textit{et al.} also reported the possible separation of a bidisperse suspension by size using the motion of a confined bubble \cite{yu2018separation}. The influence of the coupling between particles and interfacial dynamics has also recently been considered in the dip coating configuration, where a substrate is withdrawn from a particulate suspension \cite{C8SM01785A,palma2019dip}. In this situation, it has been shown that particles dispersed in a wetting fluid are entrained in the coating film when the thickness at the stagnation point is larger than a certain fraction of the particle diameter \cite{Colosqui:2013ih,sauret2019capillary,dincau2019capillary,dincau2020entrainment}. These experiments have shown that particles up to six times larger than the liquid film can contaminate the withdrawn surface \cite{sauret2019capillary}. Besides, this criterion was also found to be observed with biological microorganisms. The flow topology associated with the dip-coating configuration shares common features with the flow of a slug of liquid in a tube and the withdrawal of a liquid from a circular capillary. In particular, both configurations involve the presence of a stagnation point that governs the thickness of the coating film. Therefore, a similar condition for particle entrainment based on the particle size and the film thickness at the stagnation can be expected. To the best of our knowledge, no accurate determination of this threshold, which controls the possible entrainment of particles in a coating film on the wall of a capillary tube, has been obtained.

In this article, we report the deposition of particles during the invasion of air in a capillary tube initially filled with a dilute suspension. We predict the entrainment threshold of particles in the film and show that this situation shares common features with the dip-coating configuration, suggesting that particle entrainment is universal in the formation of thin-films governed by a stagnation point.


\section{Experimental methods}

\begin{figure}
\centering
\includegraphics[width = 0.65\textwidth]{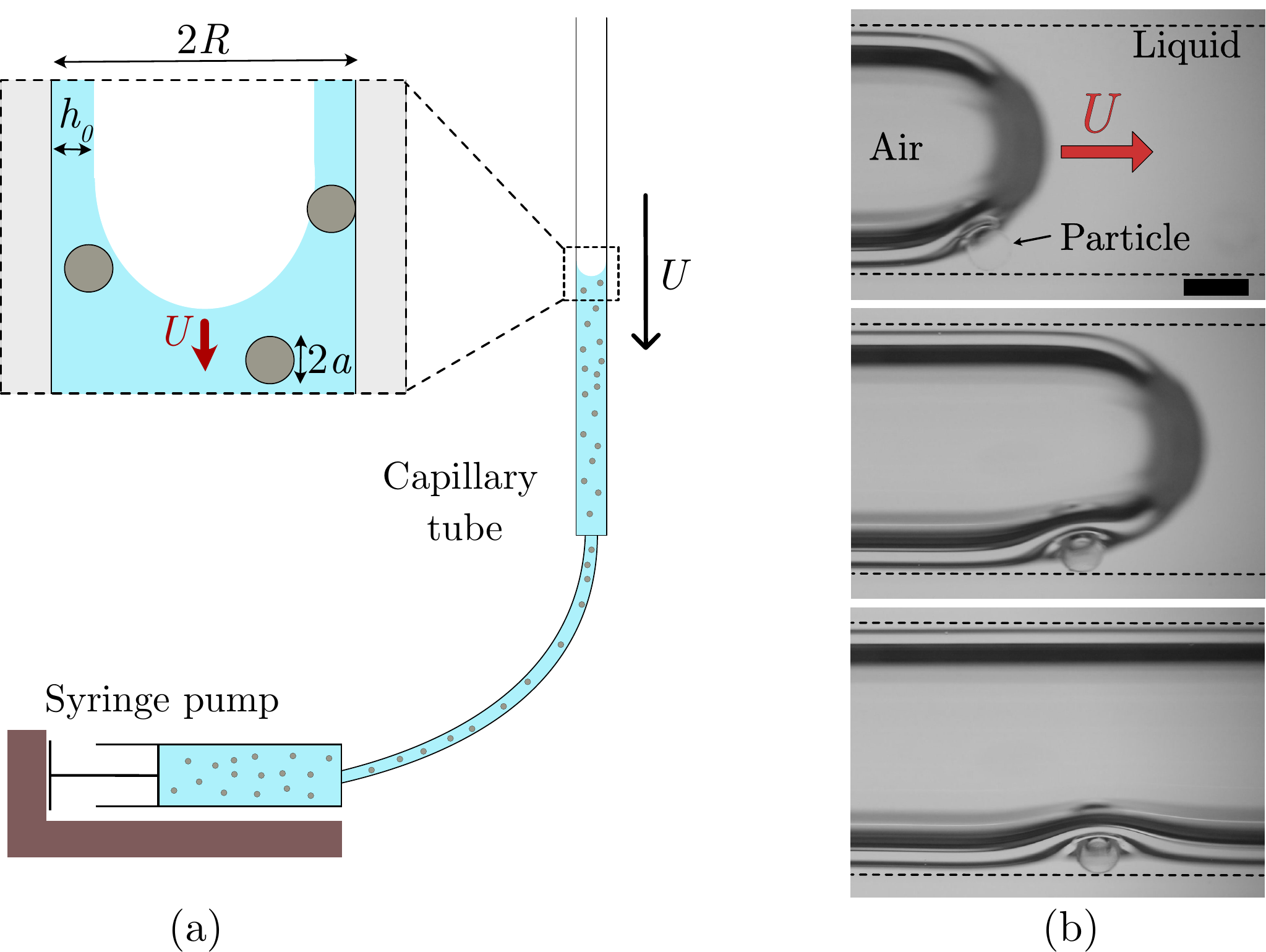}
\caption{(a) Schematic of the experimental setup. (b) Time sequence (top to bottom) of the entrainement of a particle of radius $a=125\,\mu{\rm m}$ in the liquid film for $U=4.18\,{\rm mm.s^{-1}}$ in a tube of radius $R=0.75\,{\rm mm}$ (Movie: Supplemental Material). The dashed lines indicate the inner walls of the tube. Scale bar is $400\, \mu{\rm m}$.}
\label{fig:setup}
\end{figure}

In our experiments, we initially fill a cylindrical glass capillary tube open to the atmosphere on the other side. The dilute suspension of non-Brownian particles is then withdrawn at a constant velocity $U$, leaving a liquid film of thickness $h_0$ on the inner wall of the tube [Fig. \ref{fig:setup}(a)]. The experiments are performed in capillary tubes (borosilicate glass from Vitrocom) of inner radii $R=[500,\,750,\,1000]\,\mu{\rm m}$ placed vertically. Between experiments, the capillary tubes were thoroughly cleaned with Isopropanol (IPA), acetone, and DI water and properly dried with an air gun. The suspension consists of spherical polystyrene (PS) particles (Dynoseeds TS from Microbeads) dispersed in silicone oil ($\mu=0.12\,{\rm Pa.s}$, $\gamma \simeq 22 \pm 2\,{\rm mN.m^{-1}}$), which perfectly wets the capillary tube and the particles. We used three different particle sizes and measured the size distribution of each batch through image analysis. The different particle radii used in this study are $a=40 \pm 3.5 \,{\rm \mu m}$, $a=72 \pm 4 \,{\rm \mu m}$ and $a=125 \pm 3 \,{\rm \mu m}$. The density of the particles was measured by mixing them with a mixture of DI water and Sodium Chloride (NaCl) until we reach a close density match. For the particles used here, the density is found to be $\rho_{P}=1056 \pm 2 \mathrm{kg} \mathrm{m}^{-3}$. The dilute suspensions are prepared using a precision scale (Ohaus, PX series), and the particles are dispersed in the silicone oil using a mechanical stirrer (Badger Air-Brush Paint Mixer). The density matching between the particles and the liquid allows us to consider that the particles are neutrally buoyant over the timescale of an experiment. Small volume fractions are considered, $0.32\%<\phi<1\%$, so that collective effects between particles can be neglected, and the viscosity is not significantly modified  \cite{boyer2011,guazzelli2018rheology}. The Reynolds number is small $Re \ll 1$, and inertial effects can be neglected. 

We connect one end of the capillary tube to a syringe filled with the dilute suspension. The other end of the capillary tube is open to the atmosphere. The suspension is first injected in the capillary and then withdrawn at a constant flow rate $Q$ using a syringe pump. Withdrawing the suspension, instead of injecting compressible air, allows us to reach a constant velocity of the air-liquid front quickly. The velocity of the air-liquid interface $U$ is directly related to the withdrawal flow rate through the relation $U=Q /\left(\pi\,R^{2}\right)$ and is also measured by image processing using ImageJ. To visualize the motion of the suspension, the air-liquid meniscus, and the particles deposited on the wall of the capillary, we use a backlight LED Panel (Phlox). The motion and the final morphology of the coating film are recorded with a DSLR camera (Nikon D5300) and, when needed, with a high-speed camera (Phantom VEO 710L) equipped with a macro-lens (Nikkor 200mm) and microscopic lens (Mitutoyo). The curved surface of the cylindrical glass capillary tube leads to optical distortion, and does not allow clear visualization of the liquid film and of the particles deposited on the inner wall. Therefore, we inserted the cylindrical capillary tube in a square capillary, and injected glycerol between the capillaries as its refractive index matches the index of the borosilicate glass. This method ensured that the optical distortions due to the curved walls of the inner capillary are avoided \cite{zhao2018forced}. The threshold velocity for particle entrainment in the coating film is determined as the average of the largest value of the air/liquid front velocity where no isolated particles are visible in the film and the smaller value of the air/liquid front velocity where individual particles are entrained. 

An example of the experiment is shown in Fig. \ref{fig:setup}(b), where air invades the capillary tube in the presence of particles. We observe that (i) the particle reaches the meniscus, (ii) deforms the liquid/air interface, and (iii) is entrained in the coating film whereas its diameter is larger than the film thickness. After being deposited, the particle strongly deforms the film and does not move anymore, coating the inner wall of the tube. This example shows that particle of diameter larger than the film thickness can be entrained contrary to the situation of two rigid boundaries where sieving effects would occur \cite{sauret2014clogging,dressaire2017clogging,sauret2018growth}.


\section{Results}

\subsection{Coating of Newtonian fluids}

\begin{figure}
 \begin{center}
 \includegraphics[width= 0.55\textwidth]{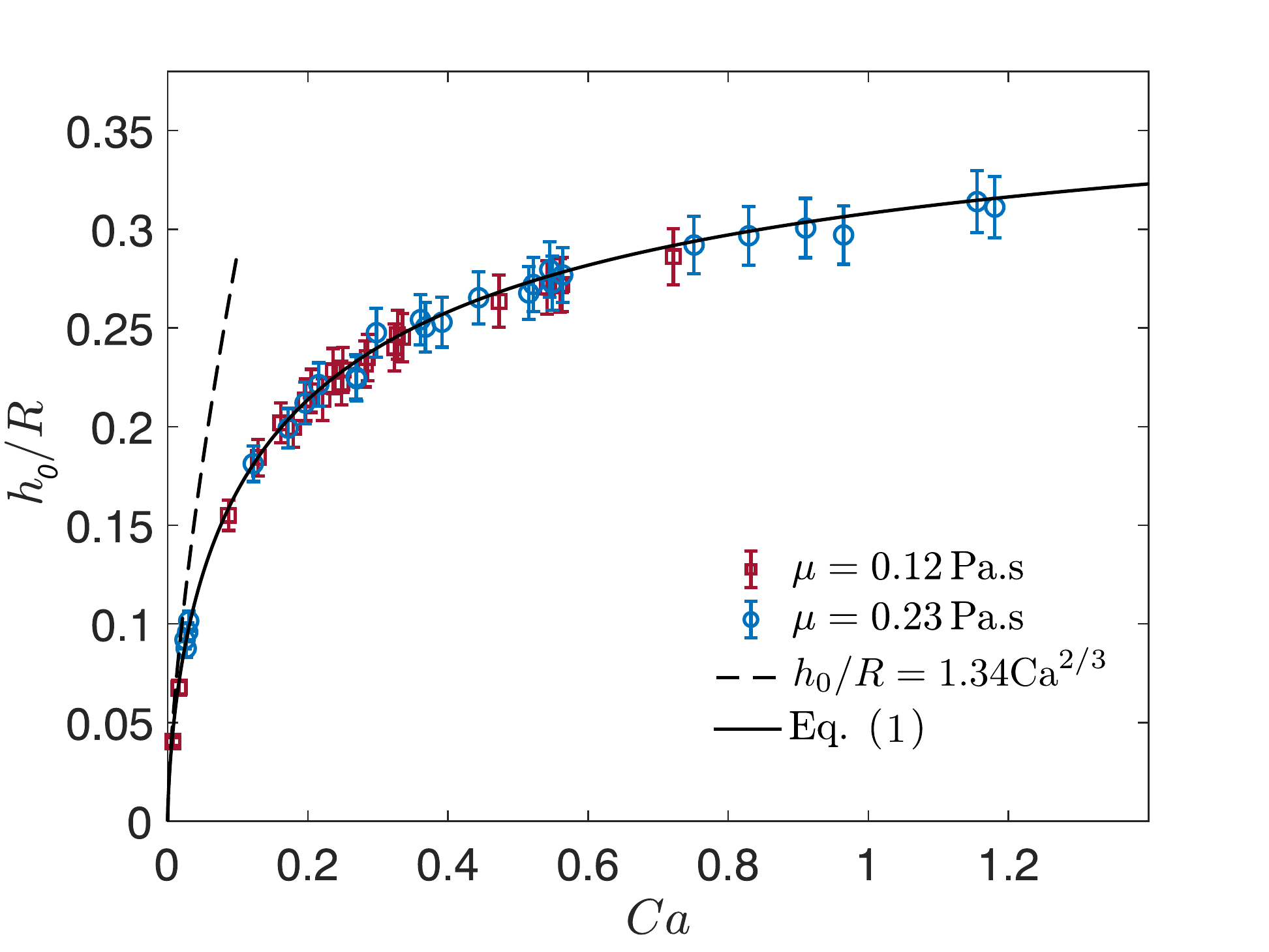}%
 \caption{Rescaled thickness $h_0/R$ of the coated film on the wall of the capillary tube when varying the capillary number ${\rm Ca}$ for $\mu = 0.12$ Pa.s (blue symbols) and $\mu = 0.24\,{\rm Pa.s}$ (red symbols). The dashed line and the solid line represent the theoretical expression ${h_0}/{R} = 1.34 {\rm Ca}^{2/3}$ and the empirical Taylor's law given by Eq. (\ref{Taylor}), respectively. \label{newtonian_fluids}}
 \end{center}
 \end{figure}

We first performed the experiments without particles for two silicone oils of different viscosities in a capillary tube of radius $R = 750\,\mu{\rm m}$. The experimental data, reported in figure \ref{newtonian_fluids}, are captured by the Taylor's law [Eq. (\ref{Taylor})] in the entire range of capillary numbers considered here \cite{balestra2018viscous,Aussillous2000}. Besides, the prediction given by ${h_0}/{R} = 1.34 {\rm Ca}^{2/3}$, valid in the limit of small capillary numbers ${\rm Ca}=\mu\, U / \gamma \ll 1$, agrees with the Taylor's law, within around 10\%, for ${\rm Ca} \lesssim 10^{-2}$. We shall see later that for the experimental parameters considered here the entrainment of particles dispersed in the liquid occurs for capillary numbers ${\rm Ca} \lesssim 10^{-2}$, thus the Bretherton's law, ${h_0}/{R} = 1.34 {\rm Ca}^{2/3}$, is expected to give a good prediction and will be used in the following. 

\subsection{Deposition threshold of particles in the coating film}

 \begin{figure}
    \begin{center}
    \includegraphics[width=0.85\textwidth]{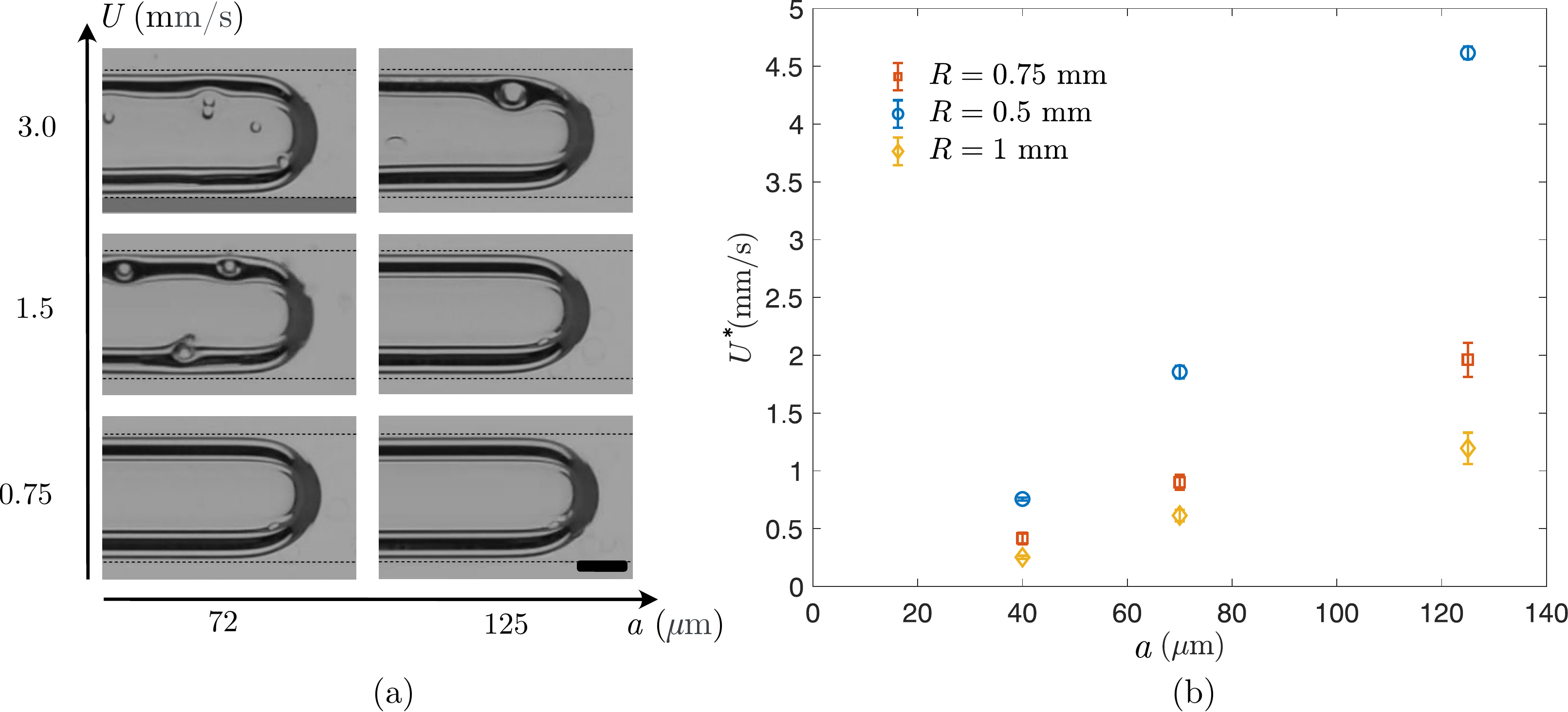}
    \caption{(a) Examples of film and particles deposited for increasing velocity (bottom to top) and two particle radii, 72 ${\rm \mu m}$ and 125 ${\rm \mu m}$, in a  tube of radius $R=750\,\mu{\rm m}$. Scale bar is 500 ${\rm \mu m}$. (b) Threshold front velocity $U^*$ beyond which particles are entrained in the coating film as a function of the particle radius $a$ for tube of radii $R =0.5,\, 0.75$, and $1 {\rm mm}$.  
    \label{entrainment_threshold}}
    \end{center}
\end{figure}

The no-deposition and deposition regimes are illustrated in Fig. \ref{entrainment_threshold}(a). At low front velocity $U$, no particles are visible in the film. Particles start to be entrained in the film beyond a threshold velocity $U^*$. The particles are randomly dispersed, and the number of particles per surface area increases with the velocity. The threshold velocity $U^*$ depends both on the radius of the particles $a$ and the radius of the tube $R$ [Fig. \ref{entrainment_threshold}(b)]. The situation observed here is reminiscent of the results reported for dip coating \cite{sauret2019capillary}. Besides, for every configuration considered here, particles are able to squeeze into films thinner than the particle diameter.

\begin{figure}
\begin{center}
 \includegraphics[width=\textwidth]{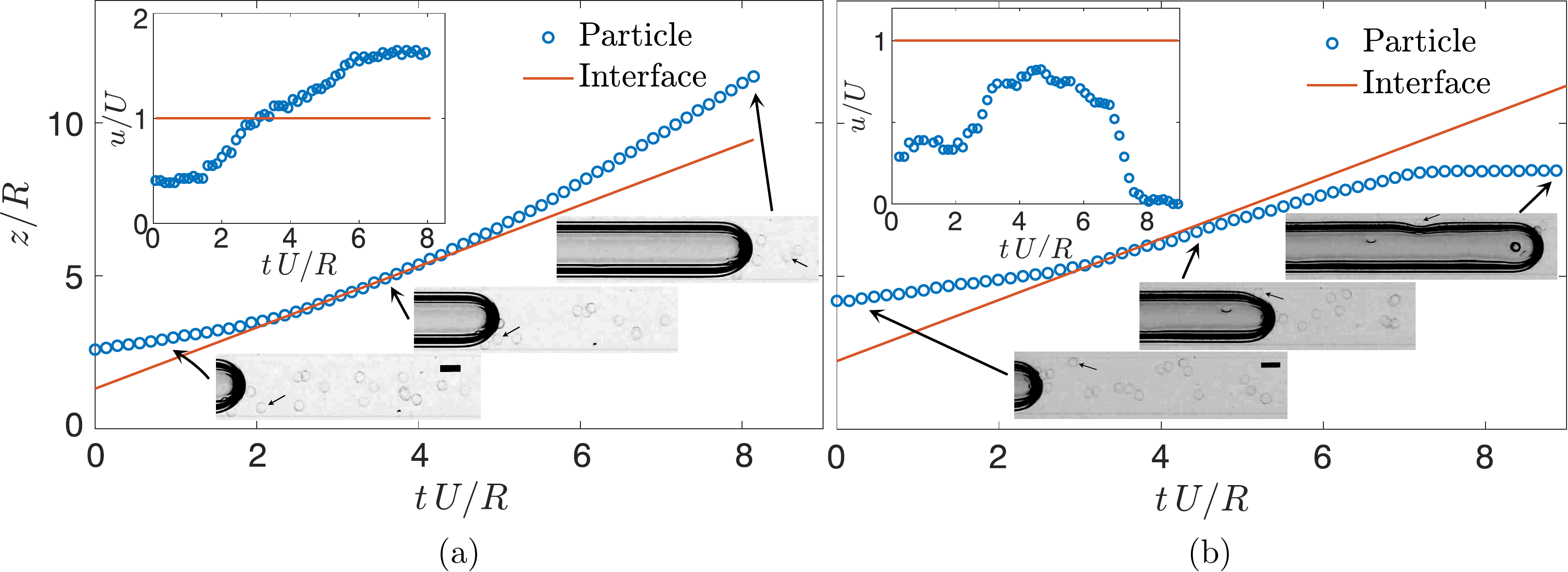}
 \caption{Particle dynamics ($R=0.75\,{\rm mm}$, $a=125\,\mu{\rm m}$) in the (a) no deposition regime for $U=2.07\,{\rm mm.s^{-1}}$ and (b) deposition regime for $U=3.45\,{\rm mm.s^{-1}}$. The main panels show the time evolution of the position of the front (solid line) and one particle (circles). Top right insets report the evolution of the rescaled absolute velocity of the front (solid line) and the particle (circles). The bottom right insets show the corresponding experimental observations. The small arrows indicate the particle and the scale bars are $500\,\mu{\rm m}$  (Movies:  Supplemental Material)
 \label{New_Figure_Bretherton}}
 \end{center}
 \end{figure}

 Examples of the time-evolution of the position of a particle are reported in Fig. \ref{New_Figure_Bretherton}(a) and \ref{New_Figure_Bretherton}(b) for the no-deposition and deposition regimes, respectively. When a particle is not entrained in the film, it first approaches the interface before recirculating in the bulk [inset of Fig. \ref{New_Figure_Bretherton}(a)]. In this regime, all particles follow similar dynamics. The entrainment case exhibits different dynamics, as illustrated in Fig. \ref{New_Figure_Bretherton}(b). The particle first approaches the interface in a similar way, but the velocity of the particle remains smaller than the front velocity. The particle is then entrained in the coating film where it reaches a zero velocity in the frame of reference of the laboratory, \textit{i.e.}, and the particle is deposited on the inner wall of the capillary [inset of Fig. \ref{New_Figure_Bretherton}(b)]. Once deposited on the tube inner wall, the particle is trapped between the wall and the air/liquid interface so that it does not move anymore and remains deposited.

 These two distinct behaviors suggest that the presence of a stagnation point on the air/liquid front is a key parameter. Indeed, the streamlines ending at this location separate the region where the fluid flows into the film and a recirculation region within the bulk. Therefore, we expect that the thickness $h^*$ at the stagnation point is the relevant parameter that controls the entrainment of particles in the film \cite[][]{Colosqui:2013ih,sauret2019capillary}. 

\section{Discussion}

\subsection{Theoretical thickness at the stagnation point}

 We consider the air invasion in a cylindrical tube in the limit ${\rm Ca} \ll 1$ leading to the deposition of a film of uniform thickness $h_0$ on the wall. The thickness $h_0$ also corresponds to the thickness observed far from the front and rear menisci during the translation of a long bubble \cite{bretherton1961motion}. In this limit, the lengthscale $\ell$, associated with the meniscus, and $h_0$, associated with the liquid film thickness far from the meniscus, satisfy $h_{0} / \ell={\rm Ca}^{1/3} \ll 1$ and $|\mathrm{d} h / \mathrm{d} x| \ll 1$ \cite{bretherton1961motion}. Besides, $h \ll R$, so that locally the situation is 2D and we use the cartesian coordinates system ($x$, $y$), where $x$ is along the capillary and $y$ is oriented inward. The interface is given by $h(x,t)$, and the front is translating at the velocity $\mathbf{U}=U\,{\mathbf{e_x}}$. The Reynolds number remains small and we use the Stokes’ equations in the lubrication analysis:
\begin{subequations} \label{eq:Stokes}
\begin{eqnarray}
u_{x}+v_{y} & = & 0 \\
0 & =& -p_{x}+\mu\,u_{y y} \\
0 & =& -p_{y}
\end{eqnarray}
\end{subequations}

  On the wall of the capillary tube, the velocity field satisfies no-slip boundary conditions, so that in the reference frame moving with the air/liquid front $u(x, 0, t)=-U$ and $v(x, 0, t)=0$. The boundary conditions at the free surface at the leading order in ${\rm Ca}$ are $\left.{\partial u}/{\partial y}\right|_{y=h}=0$ and $p=-\gamma\,\kappa$, where $\kappa=\partial_{xx}h+1/R \simeq \partial_{xx}h$. Equation \ref{eq:Stokes}(c) shows that the pressure field is only a function of $x$. Integrating equation \ref{eq:Stokes}(b) twice with respect to $y$, and using the boundary conditions leads to:
\begin{equation} \label{eq:velocity}
u(x, y)=-\frac{\gamma}{\mu} \partial_{xxx}h\left(\frac{y^2}{2}-h\,y\right)-U,
\end{equation}
To simplify this expression, we consider the classical lubrication equation \cite[see \textit{e.g.},][]{eggers2015singularities}
\begin{equation}
h_{t}+\frac{\gamma}{3 \mu}\partial_{x}\left(h^{3}\, \partial_{xxx}h\right)=0.
\end{equation}
In the reference frame moving with the front at the velocity $U$, the solution is of the form $h(x, t)=h(x-U t)$. The previous equation can be integrated with respect to $x$ together with the boundary conditions $h(x\to -\infty)=h_0$ and $\partial_{xxx}h(x\to -\infty)=0$ so that
\begin{equation} \label{eq:LLDB}
h^{2} \,\partial_{x x x}h=\frac{3\,\mu\,U}{\gamma}\,\left(1-\frac{h_0}{h}\right),
\end{equation}
which corresponds to the canonical Landau–Levich–Derjaguin–Bretherton equation. The velocity of the interface is obtained using Eq. (\ref{eq:velocity}) and Eq. (\ref{eq:LLDB}) and by setting $y=h(x)$:
\begin{equation} \label{eq:velocity_surface}
u_s(x)=u(x, y=h)=U\,\left[\frac{3}{2}\left(1-\frac{h_0}{h}\right)-1\right].
\end{equation}
The surface velocity changes sign along the interface at least once since far from the meniscus, \textit{i.e.}, $x \to - \infty$, $h(x)=h_0$ so that $u_s=-U<0$, and for $x=0$, \textit{i.e.}, at the meniscus, $h_0/h \sim 0$ so that $u_s=U/2 >0$. The thickness at the stagnation point $h^*$ corresponds to the point where the surface velocity vanishes, leading to
\begin{equation}\label{eq:stagn}
h^*=C\,h_0,\quad {\rm where } \quad C=3.
\end{equation}
This result is similar to the dip coating of a plate \cite{sauret2019capillary} and a fiber \cite{dincau2020entrainment}. Note that this analytical results applies, \textit{a priori}, only when the Bretherton law is valid, \textit{i.e.}, ${\rm Ca} \ll 1$ and $ h\ll R$. 

\subsection{Numerical simulation}

We performed numerical simulations of the fluid flow to characterize the prefactor $C$ in the range of ${\rm Ca}$, where particles are experimentally entrained. The model relies on the method developed by Balestra \textit{et al.} \cite{balestra2018viscous} and uses a laminar two phase-flow approach with moving mesh, where the fluid-fluid interface is resolved in each step by an Arbitrary Lagrangian-Eulerian (ALE) method. The initial velocity of the outer liquid is imposed as $U_{\infty}$. $U_d$ then denotes the average velocity on the surface of the droplet. The capillary number associated with the droplet is given by ${\rm Ca} = {U_d \, \mu}/{\gamma}$, where $\gamma$ is the surface tension at the interface of the droplet and ${\mu}$ is the viscosity of the surrounding liquid. The ratio between the inner and outer viscosity, $\lambda = {\mu_{in}}/{\mu}$, was fixed at 0.01 since it has been shown that this value of $\lambda$ gives the same flow profile and film thickness as those obtained at smaller values of $\lambda$ \cite{balestra2018viscous}. The ratio of densities does not affect the results and is thus set to 1. The model converges to a stationary solution when the droplet reaches its equilibrium shape and moves with a steady velocity $U_d$. The moving mesh helps to avoid large deformations of the mesh associated with the moving droplet. Furthermore, the problem is solved in the moving frame of reference attached to the droplet, the droplet hence practically does not change its position. The inlet in the model is imposed as a laminar Poiseuille flow, with mean velocity $U_{\infty} - U_d$. The velocity at the walls is imposed as $-U_d$, and the velocity of the bubble is determined and recalculated at each time step. We benchmarked the code by ensuring that the numerical simulation recovered the experimental results of Aussillous \& Qu\'er\'e for $Ca \leq 10^{-1}$ \cite{Aussillous2000}. More details on the numerical procedure can be found in Ref. \cite{balestra2018viscous}.

 \begin{figure}
\begin{center}
 \includegraphics[width= 0.85\textwidth]{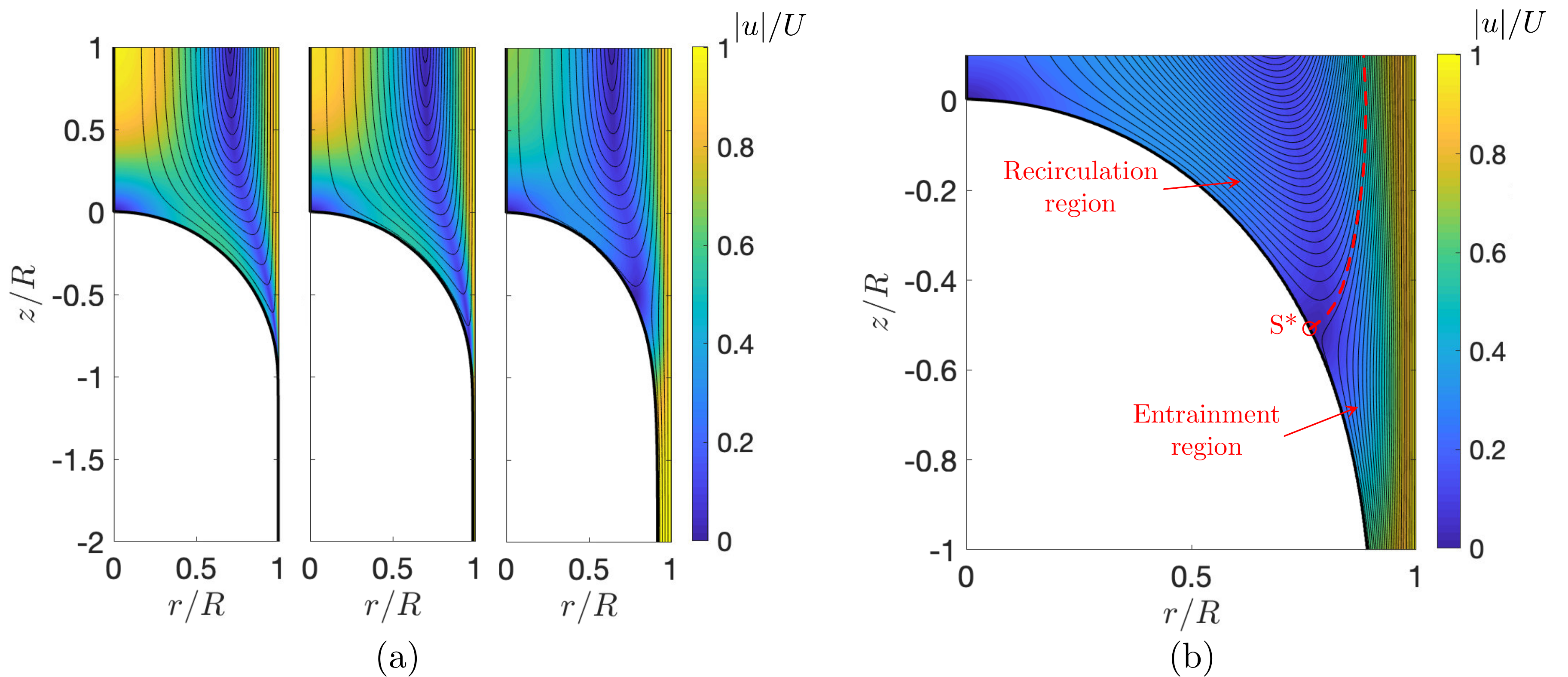}
 \caption{(a) Examples of streamlines and flow field (colormap: $\lvert u\rvert/U$) obtained numerically for ${\rm Ca}=10^{-4}$, ${\rm Ca}=10^{-3}$ and ${\rm Ca}=2 \times 10^{-2}$ (from left to right). (b) Zoom on the stagnation point $S^*$. The red dashed line separate the recirculating region where the fluid flow back to the bulk and the film region where the liquid is entrained in the coating film. \label{SM3}}
 \end{center}
 \end{figure}

Examples of streamlines observed numerically are shown in Fig. \ref{SM3}(a) for various capillary numbers. The numerical results report that the main features of the flow remain similar, but the thickness of the liquid film on the wall of the capillary increases when increasing the capillary number. Fig. \ref{SM3}(b) shows a zoom in the region surrounding the stagnation point $S^*$ that delimitates the region where the fluid flows into the film and a recirculation region within the bulk.

\begin{figure}
    \begin{center}
    \includegraphics[width=0.45\textwidth]{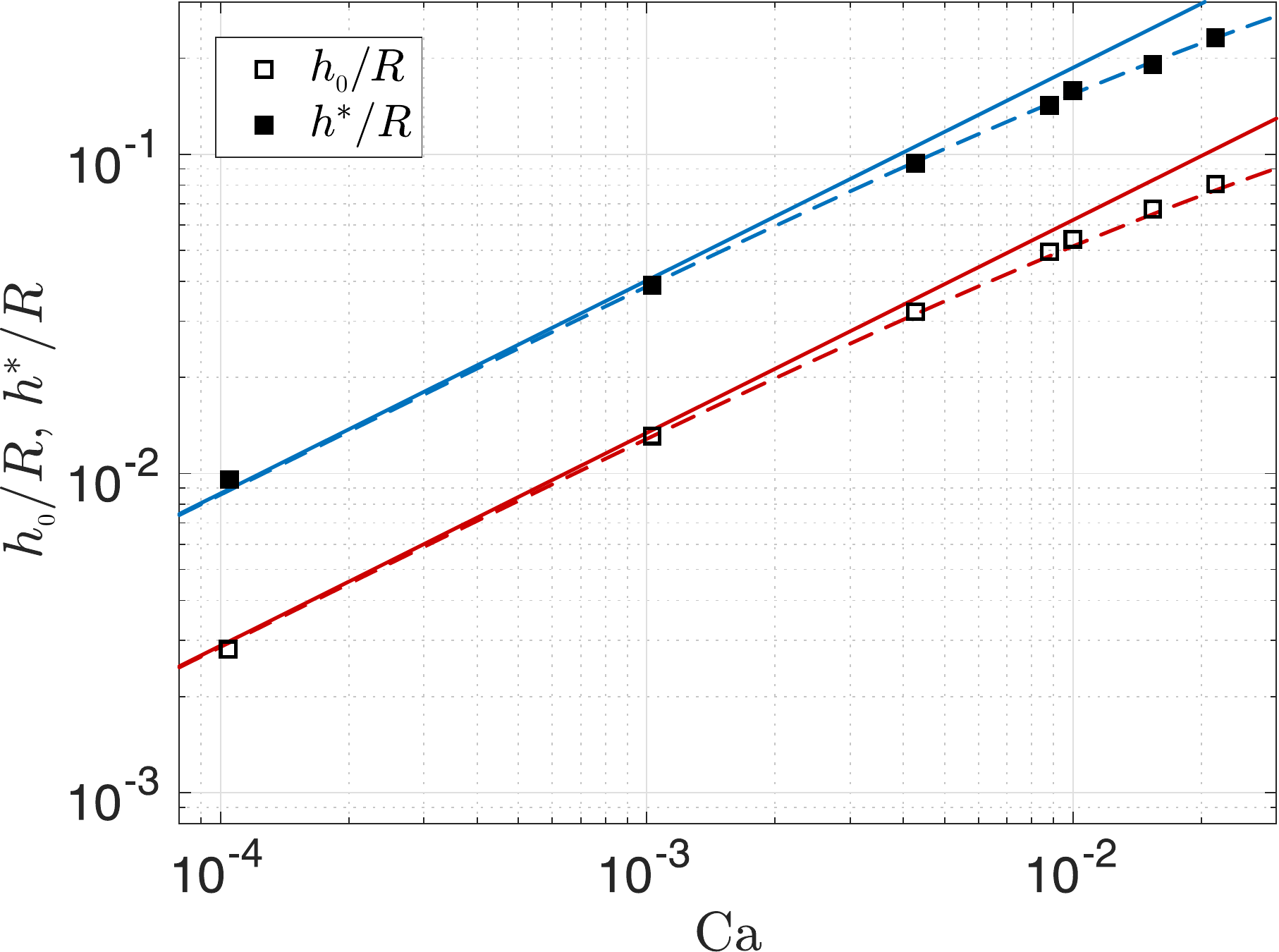}
    \caption{Evolution of $h_0/R$ and $h^*/R$ with ${\rm Ca}$. The squares are the numerical results, the solid red line is ${h_0}/{R} = 1.34 {\rm Ca}^{2/3}$, and the solid blue line corresponds to $h^*/R=3\,h_0/R$. The red dashed line is given by Eq. (\ref{Taylor}), and the blue dashed line by Eq. (\ref{stagnation_Taylor}).}
    \label{Numeric}
    \end{center}
\end{figure}

We measure the thickness $h_0$ in the uniform film region. The stagnation point corresponds to the location at the interface where the surface velocity vanishes allowing us to measure $h^*$. The dimensionless thicknesses $h_0/R$ and $h^*/R$ are reported in Fig. \ref{Numeric}. The simulations agree with the theoretical expression (\ref{eq:stagn}). Furthermore, the numerical results demonstrate that the coefficient $C=3$ obtained analytically can also be used with the Taylor's law given by Eq. (\ref{Taylor}) for a better prediction of $h^*$ for $Ca \sim 10^{-2}$ so that:
 \begin{equation}\label{stagnation_Taylor}
 \frac{h^*}{R} = \frac{4.02\, {\rm Ca}^{2/3}}{1+3.35\,{\rm Ca}^{2/3}}.
 \end{equation}

\subsection{Entrainment threshold}

\begin{figure}
    \begin{center}
    \includegraphics[width=\textwidth]{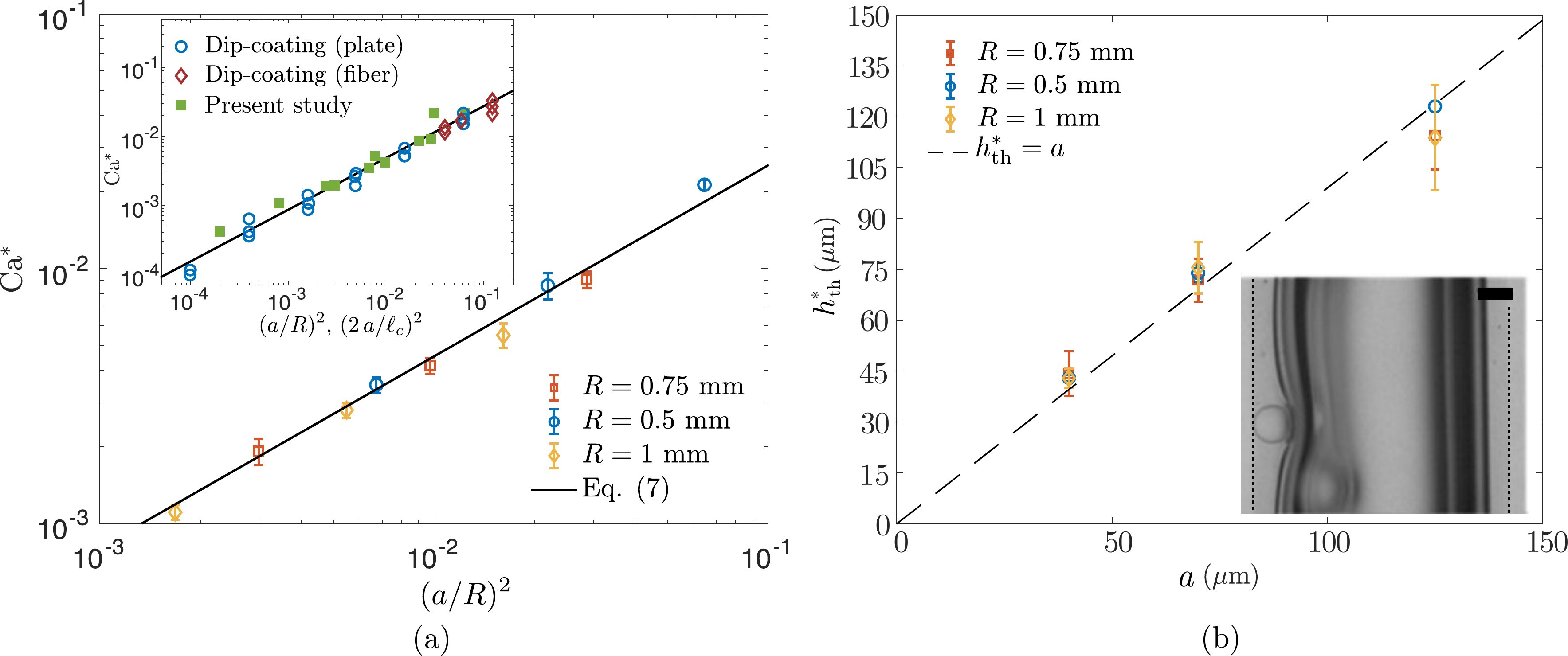}
    \caption{(a) Threshold capillary number ${\rm Ca}^*$ as a function of $(a/R)^2$ for different radii of the particles and of the tubes. Inset: Comparison with the entrainment thresholds obtained for the dip coating of a plate \cite{sauret2019capillary}, a thin fiber \cite{dincau2020entrainment} and the present experiments. In both figures the solid line is Eq. (\ref{eq:Condition2}) with $\alpha =1.15$. (b) Thickness at the stagnation point calculated theoretically (knowing the experimental parameters) allowing a particle to be entrained in the coating film as a function of the radius of the particle $a$. The dashed line shows the linear law: $h^*_{\rm th}=a$. Inset: Entrainement of a $125 \,{\rm \mu m}$ particle in the coating film (${\rm Ca} = 6.1 \,\times10^{-2}$) illustrating that an entrained particle can be smaller than the liquid film thickness owing to the deformation of the air/liquid interface. The scale bar is 200 $\mu {\rm m}$.}
    \label{Ca_lim}
    \end{center}
\end{figure}

Fig. \ref{Ca_lim}(a) reports that the threshold capillary number ${\rm Ca}^{*}$ for particle entrainment increases with $(a/R)^2$. The larger the particle is, the larger the coating film must be, and thus the associated capillary number, to entrain the particle. A particle is entrained if the thickness at the stagnation point is larger than a fraction of the particle diameter $2 a$ leading to the condition $h^{*} \geq \alpha \,a$, where $1 \leq \alpha \leq 2$. The particle can be entrained only if it follows the streamlines entering the coating film, corresponding to the condition $\alpha_{\rm min} =1$. Besides, if the particle does not deform the air/liquid interface, no capillary force is exerted on the particle, which will always be entrained in the film, so that an upper bound is $\alpha_{\rm max} =2$. Assuming that an isolated particle does not significantly modify the flow topology and using Eq. (\ref{stagnation_Taylor}), we obtain the capillary threshold value for particle entrainment:
\begin{equation} \label{eq:Condition2}
\mathrm{Ca}^{*}=\left(\frac{\alpha}{4.02}\right)^{3 / 2}\,\left(\frac{a}{R}\right)^{3/2}.
\end{equation}
Fig. \ref{Ca_lim}(a) shows that all the experiments performed collapse on a master curve given by Eq. (\ref{eq:Condition2}), where $\alpha =1.15 \pm 0.1$. The coupling of the liquid/air interface and the particle is complex. However, we can provide some rationalization of the value of $\alpha$ by considering the force acting on the particle near the stagnation point that controls its entrainment into the coating film.

\subsection{Force acting on the particle near the stagnation point}

The entrainment of the particle is initially controlled by the passive advection of the particle that follows the streamlines. Once the particle approaches the stagnation point, the experimental observations show that the particle strongly deforms the air/liquid interface. The particle is assumed to be in contact with the substrate, owing to its surface roughness of order 100 nm \cite{deboeuf2009particle,lubbers2014dense}. Therefore, the entrainment of a particle is governed by the competition between the interfacial force, the viscous drag, and the friction on the substrate. The three phases situation (particle, air, liquid) leads to a complex interplay, and we try here to estimate the forces acting on the particle. The experimental results show that the entrainment threshold occurs for a thickness at the stagnation point $h^*$ of the same order as the particle radius $a$ [Fig. \ref{Ca_lim}(b)]. Therefore, we consider the force acting on a particle of radius $a$ located at the stagnation point of thickness $h^* \sim a$. We further assume that far from the particle, the interface has a shape similar to the situation without particle.

\begin{figure}
\begin{center}
 \includegraphics[width= 0.7\textwidth]{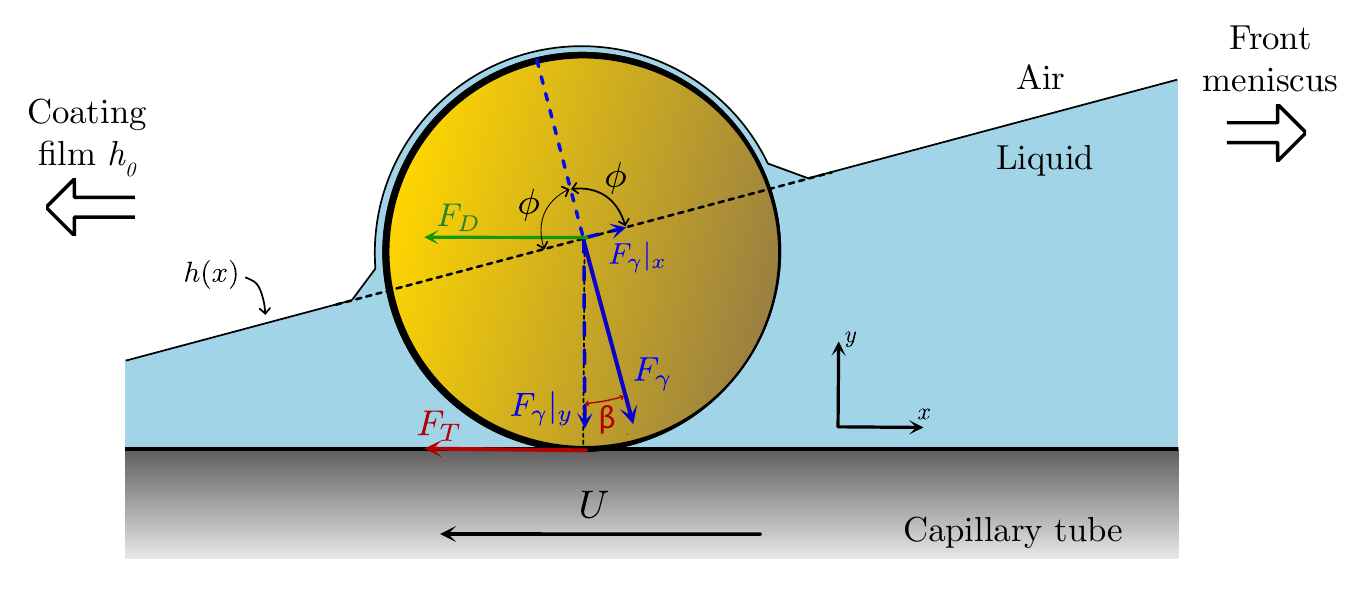}%
 \caption{Schematic of the forces exerted on a particle confined in the liquid at the stagnation point.  \label{SM6}}
 \end{center}
 \end{figure}

A schematic of a particle fully wetted by a liquid of thickness $h^*$ around the stagnation point is shown in Fig. \ref{SM6}. Below, we detail the amplitude and orientation of the different forces to estimate their influence on particle entrainment. We consider the situation in the frame of reference moving with the front. Besides, close to the entrainment threshold, the velocity of the particle nearly vanished at the stagnation point before being able to be entrained in the coating film. The experimental parameters imposed that inertial effects are negligible since ${\rm Re} \ll 1$.

An estimate of the viscous drag acting on the particle can be obtained by considering the average viscous stress acting on the particle: $\mu \,U / (2\,h^*)$ \cite{lubbers2014dense}. Furthermore, the viscous stress acts over half of the particle since $h^* \sim a$ so that the total drag is of order
\begin{equation} \label{eq_drag}
F_D \sim \frac{\mu \,U}{2\,h^*}\,\frac{\pi\,a^2}{2} \sim \frac{\mu \,\pi\,U\,a}{4}.
\end{equation}
The drag force acts along the $x$-axis and acts as a driving force to entrain the particle (see Fig. \ref{SM6}).

\medskip

Since the particle is totally wetted by the liquid (contact angle: $\theta=0$), the capillary force is acting downard, pushing the particle against the wall of the capillary tube. The amplitude of the capillary force is given by (see, \textit{e.g.}, \cite{gomez2001analysis}):
\begin{equation}
F_{\gamma}=2 \,\pi\,a\, \gamma \sin \phi \sin (\theta-\phi),
\end{equation}
where $a$ is the radius of the particle, $\gamma$ the interfacial tension, $\phi$ the filling angle and here $\theta=0$ since the silicone oil perfectly wets the particle. For a liquid film of thickness $h \sim a$, the absolute value of the amplitude of the capillary force reduced to $F_{\gamma}=2 \,\pi\,a\, \gamma \sin^2 \phi$. The capillary force is maximum for $\phi=\pi/2$, corresponding to $h \sim a$, so that an order of magnitude of the capillary force acting on the particle is 
\begin{equation}
F_{\gamma} \sim 2 \,\pi\,a\, \gamma
\end{equation}
As illustrated in Fig. \ref{SM6} the capillary force can be decomposed in its $y$ component that pushes the particle toward the surface of the capillary tube
\begin{equation} \label{Fcapy}
F_{\gamma}\rvert_y \sim - 2 \,\pi\,a\, \gamma \cos{\beta}
\end{equation}
and an axial force acting to keep the particle in the liquid reservoir (in the direction of the front):
\begin{equation} \label{Fcapx}
F_{\gamma}\rvert_x \sim 2 \,\pi\,a\, \gamma \sin{\beta}
\end{equation}
The angle $\beta$ comes from the inclination of the air/liquid interface between the front and the region of uniform coating thickness. Note that we have neglected the radial curvature of the capillary since both $h^*$ and $a$ are very small compared to $R$. The slope at the interface close to the stagnation point is obtained by linearizing the Landau-Levich-Derjaguin-Bretherton (LLDB) equation for the thickness \cite{Colosqui:2013ih,Landau1942}:
\begin{equation}
h(x)=h_0+\left(h^{*}-h_0\right) \exp \left(-\frac{\left(x-x^{*}\right)}{h_0/(3\,Ca^{1/3})}\right),
\end{equation}
As a result, an approximate solution of the slope is $\tan \beta={\rm d} h(x^*)/{\rm d}x \sim -3\,{\rm Ca}^{1/3}$. In the range of parameters considered here leading to the entrainment of the particles ${\rm Ca} \sim 10^{-3}-10^{-2}$, so that $\beta \sim 15^{\rm o} - 35^{\rm o}$.

\medskip The resistance of the particle motion when the front is advancing and trapping of the particle at the stagnation point occurs because the particle becomes in physical contact with the glass substrate \cite{lubbers2014dense}. The particle is thus subject to a dynamic friction $F_T=\mu_d\,F_N$, where $\mu_d \simeq 0.3$ is an estimate of the dynamic friction coefficient between the immersed polystyrene particle and the glass substrate \cite{tapia2019influence} and $F_N$ denotes the normal force confining the particle against the substrate, \textit{i.e.} the $y$ component of the capillary force given by Eq. (\ref{Fcapy}). Therefore, the dynamic friction force, oriented along $x<0$ and opposed to the motion of the front, is
\begin{equation} \label{FT}
F_T \sim  2\,\mu_d \,\pi\,a\, \gamma \cos{\beta}
\end{equation}

\medskip 

 We can now estimate the order of the magnitude of the two opposite forces induced by the deformation of the interface by the particles. The $x$-component of the capillary force tends to keep the particle within the bulk liquid, whereas the $y$-component confine the particle and leads to the friction force that favors the particle entrainment during the translation of the air/liquid front. Using Eqs. (\ref{Fcapx}) and (\ref{FT}) together with the expression of $\beta$, leads to an order of magnitude of the ratio of these forces:
\begin{equation}
\frac{F_{\gamma}\rvert_x}{F_T }=\frac{\tan \beta}{\mu_d} \sim \frac{3\,{\rm Ca}^{1/3}}{\mu_d},
\end{equation}
For the order of magnitude of capillary number leading to capillary entrainment in this study, ${\rm Ca} \sim 10^{-3}-10^{-2}$, we found that this ratio is of order unity (between 0.8 and 1.85). Therefore, the effects of the capillary force induced by the deformation of the interface mostly balance each other.

\medskip 

Using the order of magnitude of the experimental parameters used in this study, we can estimate the drag force and the friction force that also contribute to entrain the particle in the coating film and the capillary force along $x$ that prevents the deposition of the particle in the coating film. An order of magnitude of the ratio of these two opposite effects is given by

\begin{equation} 
\frac{F_D+F_T}{F_{\gamma}\rvert_x} \sim \frac{\rm{Ca}}{8}+ \frac{\mu_d}{3\,{\rm Ca}^{1/3}} \sim \frac{\mu_d}{3\,{\rm Ca}^{1/3}}.
\end{equation}
The two components of the capillary forces mainly balance each other owing to the friction force induced by the contact of the particle on the wall. The main limitation to entrain the particle in the film is thus the possibility for a particle to follow the streamlines leading to the coating flow, corresponding to the condition $\alpha \sim 1$ as observed experimentally.

The value of $\alpha$ is also in quantitative agreement with previous experiments on dip-coating of plates and thin fibers \cite{sauret2019capillary,dincau2020entrainment} as reported in the inset of Fig. \ref{Ca_lim} (note that the because of the flat geometry, the relevant parameter for the plate configuration is $\left( 2\,a/\ell_c\right)^2$ to account for the difference in geometry). Therefore, the entrainment of particles in a coating film governed by a stagnation point is controlled by a universal mechanism for dip coating, coating of the inner wall of a tube, and is expected to hold for a confined air bubble in a perfectly wetting fluid. The translation of a long bubble beyond the entrainment threshold would lead to similar results, except that particles will first be trapped in the liquid film and then released in the liquid at the rear of the bubble. In particular, contrary to the recent experiments of Yu \textit{et al.} \cite{yu2018time}, no attachment of particles at the air/water interface of the bubble would be observed because the liquid fully wets the particles.

We should emphasize that we have considered diluted suspension in this work so that particles can be considered as isolated \cite{sauret2019capillary}. For larger volume fractions of the suspension, the entrainment of clusters may be observed as well as the accumulation of particles near the meniscus. We should emphasize that we have never observed clogging of the tube for the perfectly wetting fluid considered here, but we have observed, from time to time, the entrainment of some clusters near the entrainment threshold.

\section{Conclusion}

In this article, we have shown that the liquid coating the inner wall of a capillary tube exhibits different regimes in the presence of particles: liquid only and deposition of particles. We demonstrated that a particle is entrained in the film even if its diameter is larger than the film thickness. Considering the thickness at the stagnation point, the experimental results have been quantitatively rationalized. In particular, we have provided some physical insights to rationalize the order of magnitude of the parameter $\alpha$. A similar physical mechanism underlies the entrainment of particles as observed in both this configuration and in dip coating \cite{Colosqui:2013ih,sauret2019capillary}. This result comes from the common features between this configuration and the dip-coating of a substrate, the most important of which is the presence of a stagnation point. Therefore, preventing the contamination of substrate by particles and biological microorganisms using this strategy, as reported for the dip coating system \cite{sauret2019capillary}, is also valid in tubings and can provide guidelines for dispensing of suspensions and contaminated liquid in tubes. These results are also relevant to environmental processes involving the transport of small particles in confined geometries, such as the dissemination of colloid and microorganisms in porous media \cite{saiers2003role} and provide guidelines to improve the coating of tubings by suspensions \cite{primkulov2020spin}. This filtering mechanism could also be an ingredient to explain the increase of bubble rise speed observed recently in confined suspension \cite{madec2020puzzling}. The influence of the wettability of the particle can further refine the value of the prefactor and is the topic of an ongoing study.

\begin{acknowledgments}
This material is based upon work supported by the National Science Foundation under NSF CAREER Program Award CBET No. 1944844 and by the ACS Petroleum Research Fund 60108-DNI9. We thank F. Gallaire and L. Keiser for their help with the numerical model and B. Dincau for a careful reading of the manuscript.
\end{acknowledgments}

\bibliography{BiblioBretherton}

\end{document}